# Half-integer Topological Defects Paired via String Micelles in Polar Liquids

Zhongjie Ma[1], Miao Jiang[1*], Yaohao Song[2], Aile Sun[1], Shengzhu Yi[1], Chao Zhou[1], Xiang Huang[2], Mingjun Huang[2,3], Satoshi Aya[2,3], Qi-Huo Wei[1,4*]

1 Department of Mechanical and Energy Engineering, Southern University of Science and Technology; Shenzhen, 518055, China.

2 South China Advanced Institute for Soft Matter Science and Technology (AISMST), School of Emergent Soft Matter, South China University of Technology; Guangzhou 510640, China.

3 Guangdong Provincial Key Laboratory of Functional and Intelligent Hybrid Materials and Devices, South China University of Technology; Guangzhou 510640, China.

4 Center for Complex Flows and Soft Matter Research, Southern University of Science and Technology; Shenzhen 518055, China.

*Qi-Huo Wei **Email:** weiqh@sustech.edu.cn

*Miao Jiang **Email:** jiangm@sustech.edu.cn

**Abstract**

Ferroelectric nematic ($N_F$) liquid crystals present a compelling platform for exploring topological defects in polar fields, while their structural properties can be significantly altered by ionic doping. In this study, we demonstrate that doping the ferroelectric nematic material RM734 with cationic polymers enable the formation of polymeric micelles that connect pairs of half-integer topological defects. Polarizing optical microscopy reveals that these string defects exhibit butterfly textures, featured with a two-dimensional polarization field divided by Néel-type kink-walls into domains exhibiting either uniform polarization or negative splay and bend deformations. Through analysis of electrophoretic motion and direct measurements of polarization divergences, we show that the string micelles are positively charged and their side regions exhibit positive bound charges. To elucidate these observations, we propose a charge double layer model for the string defects: the positive charged cationic polymer chains and densely packed RM734 molecules form a Stern charge layer, while small anionic ions and positive bound charges constitute the charge diffusion layer. Notably, our experiments indicate that only cationic polymer doping effectively induces the formation of these unique string defects. These findings enhance our understanding of ionic doping effects and provide valuable insights for engineering polar topologies in liquid crystal systems.

## Introduction

Topological defects are ubiquitous in condensed matter systems with broken continuous symmetry, playing a crucial role in mediating phase transitions and dictating material properties (1-4). Two exemplary states of matter that demonstrate such symmetry breaking are the nematic (N) liquid crystal (5) and the recently discovered ferroelectric nematic ($N_F$) liquid crystal (6-9). In the ground state of nematic liquid crystals, rod-shaped molecules align in a common direction without positional order. This averaged molecular orientation, referred to as the director **n**, serves as the order parameter (10). The director **n** is a headless vector, adhering to inversion symmetry, **n**(r)≡ -**n**(r). The space for the degenerate values of director **n** is represented by a circle in two dimensions (2D) or a spherical surface in three-dimensions (3D), with diametrically opposite points identified (3, 10, 11). Nematic liquid crystals have been a fertile platform for exploring the intricate physics (12, 13) and applications of topological defects such as programable origami (14), command of active matter (15), molecular self-assembly (16) and topological light (17). In contrast, in $N_F$ liquid crystals, the order parameter is the spontaneous polarization ***P*** (7, 18), whose polar nature disrupts inversion symmetry. The order parameter space for ***P*** forms a circle in 2D and a spherical surface in 3D (18, 19), necessitating that topologically stable defects exhibit integer winding numbers. The winding number is defined as the number of rotations that the polarization vector ***P*** makes in the order parameter space when encircling a singularity in real space.

The ferroic materials with the polarization confined to a 2D plane are often represented by the classical XY model (20). A theory based on the XY model predicts the excitations of vortex and antivortex pairs with half integer winding numbers. These half-integer defect pairs are connected by string defects with a line tension that may diminish with temperature, leading to a phase transition (21). There has been considerable interest in exploring half-integer vortices and strings in systems with similar order parameter such as tilted smectic liquid crystals (22), liquid $^3$He (23) and Bose–Einstein condensate (24). To the best of our knowledge, direct observations of the string defects paired with half-integer vortices in polar materials have not been reported.

Electrical screening can modify the electrostatic environment within ferroelectric materials, alleviating the effects of depolarization fields caused by bound charges at interface or within the material itself (25). Doping the ferroelectric materials with molecules carrying various charges is an important avenue to enhancing or weakening the electrical screening effects (26). In $N_F$ liquid crystals, the large spontaneous polarization leads to polydomain structures or chiral ground state to reduce electrostatic energy (9, 26). It has been observed that even trace amounts of ionic doping can significantly modify the electric properties of $N_F$ material and manipulate the phase transitions (27), as well as lower the tendency to form chiral ground states (26). Significant progress has been made in developing ferroelectric nematic materials and understanding their physical properties such as structures and domain walls (18, 28-32), while our understanding on ionic doping effects is still incomplete.

In this study, we investigate the effects of electrical screening on topological defects in the ferroelectric nematic liquid crystal RM734. We demonstrate that doping with cationic polymers facilitates the formation of pairs of ±1/2 topological defects connected by polymeric micelles. Through polarizing optical microscopy, we observe butterfly textures characterized by a two-dimensional polarization field divided by Néel-type kink walls into domains exhibiting uniform polarizations alongside splay and bend deformations. By analysing the polarization field and the electrophoretic motion of string defects, we show that the micelles consist of positively charged doped polymers, while the butterfly wing regions display positive bound charges. These findings support a charge double-layer model: the polymer chain micelles and densely packed RM734 molecules form a Stern layer, while small anionic ions and the positive bound charges in the wing regions constitute the diffusion layer. We also compared various ionic dopants and found that only cationic polymers can induce the formation of string defects. This behaviour is attributed to the electrostatic interactions between the dopants and the liquid crystal molecules, as well as the flexoelectric coupling effects inherent to the pear-shaped RM734 molecules.

## Results

**Butterfly Textures and the Structures of the String Defects.** We employed the ferroelectric nematic liquid crystal material RM734 in our experiments. The RM734 molecule exhibits an electrical dipole moment of ~11 Debye and bears a pear-like asymmetric shape (7), with the electrical dipole oriented towards the large head. Pure RM734 undergoes a transition from the isotropic (I) to the N phase at 182°C, followed by a transition to the $N_F$ phase at 132°C.

We introduced a cationic polymer into the RM734, specifically poly (dimethyloctadecyl 3-trimethoxysilyl propyl ammonium chloride) (PDMOAP). Remarkably, even a trace amount of PDMOAP can induce significant reduction in the phase transition temperature. For instance, with just 0.05 wt% of PDMOAP doped in RM734, we observe a 4°C reduction in the N-$N_F$ transition temperature (SI Appendix, Fig. S1).

We assembled liquid crystal cells using cleaned glass slides without alignment agents, and filled these cells with the RM734 and PDMOAP mixture in its isotropic phase at temperatures exceeding 180°C. We then cooled the cells at a rate of 5°C/min. In the N phase, the liquid crystal molecules are aligned vertically to the surfaces, as evidenced by the dark images under a cross-polarized optical microscope and the dark cross in the conoscopic image at 140°C (Fig. 1A). Additional experiments reveal that for the pure RM734 material, the molecules are aligned parallel to the substrate surface with no preferential direction. These observations suggest that the ionic polymers PDMOAP self-assemble on the substrates and form monolayers to facilitate the vertical alignment of the liquid crystal molecules in the N phase.

When the temperature is below 128°C, the liquid crystal mixture enters the $N_F$ phase. We conducted polarization measurements and obtained an instantaneous polarization of ~5 μC/cm$^2$ for the mixture at 120°C (SI Appendix, Fig. S2). This matches the polarization value of ~6 μC/cm$^2$ for pure RM734 and verifies that the liquid crystal is in the $N_F$ phase (7).

The molecular orientations (also the polarization) in the $N_F$ phase are aligned parallel to the confining surfaces, as evidenced by the uniformly dark or bright polarizing optical images (Figs. 1B and C). Especially, numerous butterfly-like textures can be observed with a central string dividing two luminous wings and a dark line traversing through it (Fig. 1B) during the phase transition (SI Appendix, Video. S1). These butterfly wings are mirror-symmetric with respect to the string, but asymmetric along the string. We refer to the string ends with narrow and wide wings as the head and tail, respectively (Fig. 1D). When the cross-polarizers are rotated by 45º, the brightness of the texture changes from bright into dark, indicating that there is no twist of the molecular orientation across the cell (Fig. 1C). The string defects retains their configuration upon further cooling of the $N_F$ phase.

To ascertain the polarization field of the butterfly texture, we applied an in-plane electrical field and observed its response (SI Appendix, Fig. S3). When the electrical field is from the tail to the head, the butterfly textures remain unchanged and moving, showing that the polarization is pointing in the same direction (Fig. 2A). We then measured the director field by using the PolScope technique (33) and obtained the polarization fields by combining the director field with the polarization direction (Fig. 2B).

The measured polarization field displays several notable characteristics. Firstly, the polarization lies predominantly in the plane of the cell. The uniform phase retardation and no twist texture observed in cross polarized optical images indicate that the order parameter $\boldsymbol{P}$ of the system lacks a z-component, making the system analogous to a XY model. Secondly, the strings are connecting two half-integer defects at its ends. Specifically, the +1/2 defect is situated at the head, while the -1/2 defect is located at the tail (Fig. 2B). Thirdly, large variations in the polarization field are discernible at the edges of the butterfly wing textures, where discontinuities in optical intensities are visible (Fig. 2A). We plot the profiles of the orientation angle $\theta$ of the polarization along horizontal directions and observe step-like variations (Fig. 2C). These variations can be well fitted with the kink function: $\theta = \pm\theta_0 \tanh(\sqrt{2}x/2\xi) + \theta_0$ (34), where the $\pm$ sign represents the kink and anti-kink at the left and right sides of the wing, $2\theta_0$ is the angle that the polarization rotates when going across the kink. Best fitting yields the kink width $\xi$~1.8 μm for left and $\xi$~4 μm for right edges (Fig. 2C).

Such kinks are domain walls dividing the polarization fields into the regions with primarily uniform directors. These kink walls are different from the normal 180° domain walls in ferroelectric materials. Firstly, for the 180° domain walls, the polarizations at two neighbouring sides are parallel to the wall but rotated by 180°. Secondly, the 180° domain walls are typically of the Ising-type characterized by the polarization decreasing to zero, changing sign and increasing to equilibrium value, and the polarization magnitude is a one-dimensional kink function (34). While for the kink walls here, the polarization changes direction mainly via the rotation of the polarization vector around an axis parallel to the wall. This is a signature of the Néel type of walls often encountered in ferromagnetic materials (35). These kink walls were observed in two previous publications (29, 36) and called the p-walls in Ref. (29) considering that the shape of the domain wall is parabola. Variation of the bending angle along the kink wall is continuously decreased to uniform domain, indicating the boundary configuration for the kink wall is a flexible range larger than 20° rather a specific polar angle (Fig. 2C).

We calculated the bend deformation of the director field, expressed as $\boldsymbol{B}(x, y) = \boldsymbol{n} \times (\nabla \times \boldsymbol{n})$ (Fig. 2D). Due to the 2D nature of $\boldsymbol{n}(x, y)$, the bend deformation $\boldsymbol{B}$ is a vector perpendicular to the xy plane. We find that the bend deformation is maximal at the left edge of the wing (i.e. at the kink) and is finite around the right-wing edge. In the remaining director field, the bend deformation is close to zero, with a uniform director field. Further analysis shows that the peak line of $|\boldsymbol{B}|$ roughly bisect the angle formed by the polarizations ($\boldsymbol{P_1}$ and $\boldsymbol{P_2}$) beside it (Fig. 2B). This implies that the normal component of the polarization is continuous along the kink-like walls, or $(\boldsymbol{P_2} - \boldsymbol{P_1}) \cdot \boldsymbol{V} = 0$, where $\boldsymbol{V}$ is a unit vector normal to the kink. This configuration is energetically favourable, as the bound electric charge at these kink walls, expressed as $\sigma = (\boldsymbol{P_2} - \boldsymbol{P_1}) \cdot \boldsymbol{V}$, is zero.

We also calculated the splay deformation of the director field, expressed as $S(x, y) = \nabla \cdot \boldsymbol{n}(x, y)$ (Fig. 2E). For the RM734, the electrical polarization is approximately parallel to the director, $\boldsymbol{P} = P_0\boldsymbol{n}$, where $P_0$ is the equilibrium value of the spontaneous polarization. Given that the divergence of the polarization equals the bound charge density $\rho = -\nabla \cdot \boldsymbol{P}$, the splay deformation represents the distribution of the bound charge density, $S(x, y) = \rho/P_0$. The butterfly wing areas are negatively charged, while the string and its proximity regions are positively charged (Fig. 2E). A plot of $S(x, y)$ along the vertical direction (for fixed $x$) shows that the bound charge density decays exponentially with the distance $y$ from the string $S(x, y) \approx S_0 e^{-\kappa y}$, where the decay length $1/\kappa$ is between 8 to 11 μm (SI Appendix, Fig. S4).

To determine the spatial distribution of the doped ionic polymers, we replaced the PDMOAP with a Rhodamine-B-labelled cationic polymer, named poly (rhodamine-3-aminopropyl triethoxysilane) (SI Appendix, Fig. S5). At the central emission wavelength (~575 nm) of Rhodamine B, confocal fluorescence microscopic imaging reveals that significant fluorescent emissions are from the strings, indicating that the doped cationic polymers are concentrated in the strings (Fig. 2F).

To note, when the concentration of cationic polymers is sufficiently low, the strings disappear, and the liquid crystals appear as a large monodomain without domain walls (SI Appendix, Fig. S6). This is similar to the existence of the critical micelle concentration (CMC) in solutions of amphiphilic molecules (37). Below the CMC, the amphiphilic molecules are primarily dispersed as individual molecules or small aggregates, while above the CMC, the amphiphilic molecules start to form micelles of various shapes such as spherical and filamentous. Here the polymer aggregates exhibit the string configuration which can be attributed to a balance between bulk elastic energy and surface tension (38, 39).

In a commentary, Lavrentovich envisioned that two half-integer disclinations in a two-dimensional polarization field can serve as seeds and edges of domain walls, and that two topologically identical configurations are possible, one weakly charged, and the other strongly charged (as reproduced them in Fig. 2G and 2H) (19). To note, the weakly charged configuration is the same as the string defect predicted in the XY model (21).

The polarization field of the string defects are identical to the strongly charged configuration envisioned by Lavrentovich (19), and only topologically identical to prediction by the classical XY model (21). Deviations from these predictions are also obvious. First, the formation of these string defects is largely facilitated by doping of cationic polymers. The doped ions provide the electrical screening needed to neutralize the bound charges generated by the deformations of topological defects. Second, the polymer strings are three-dimensional objects, unlike domain walls. Lastly, the polarization fields near the +1/2 and -1/2 defects display noticeable asymmetry, with bending deformations concentrated in Néel-type kinks near the -1/2 defect. This configuration can reduce the total splay deformation by an increase of bend deformation(29).

**Electrophoretic Motion and Interaction between String Defects.** When an electrical field parallel to the polarization vector is applied, the string defects exhibit electrophoretic motion in the $N_F$ liquid crystal. To characterize this electrodynamic motion, we varied the electrical field strength $\boldsymbol{E}$ from 0 to 5 V/mm in increments of 1 V/mm, and measured the positions of the string defects with different lengths (Fig. 3A). We find a good linear relationship between the motion velocity $\boldsymbol{v}$ and the electric field $\boldsymbol{E}$, which can be fitted by the equation $v = \mu_{EP}(E - E_0)$. This linear relationship indicates that the field-driven string motion is due to electrophoresis. Here, the electrophoretic mobility $\mu_{EP}$ varies with the string length, while the threshold field $E_0$ remains consistently at ~0.8 V/mm regardless of the string length (Fig. 3B).

The physical origin of the threshold field can be attributed to the existence of an insulating layer on the electrodes interface, due to the surface pinning of RM734 molecules (40). Below the saturated voltage, a tiny field-induced reorientations of ***P*** are sufficient for the self-field to completely cancel the E field in ferroelectric RM734 layer(31). Thus, the electric field applied appears entirely across the high-capacity interfacial layers on the electrodes.

The viscous force experienced by a moving cylinder is proportional to its motion velocity $v$: $F_v = \zeta v$, where the viscous coefficient $\zeta$ is given by the formula $\zeta = 2\pi\eta L/(ln(L/r) - 0.72)$ (41). Here, $\eta$ is the viscosity of the surrounding liquid, $L$ and $r$ are the length and radius of the cylinder, respectively. The electrophoretic force is: $F_E = \lambda L(E - E_o)$, with $\lambda$ being the electric charge density per unit length. Under the small Reynolds number (~$10^{-6}$), the electrophoretic force is balanced by the viscous force, yielding: $v = \lambda[\ln(L/r) - 0.72](E - E_0)/2\pi\eta$, so the electrophoretic mobility is:

$$\mu_{EP} = \lambda(\ln(L/r) - 0.72)/2\pi\eta \tag{1}$$

The measured $\mu_{EP}$ is plotted as a function of the string length at different electric field strengths in Fig. 3C. A best fitting with Eq. 1 yields $\lambda$~$10^{-9}$ C/m by taking $\eta$~0.1 Pa·s and $r$~200 nm (8).

As an alternative method, the charge density can be estimated from the bound charges in the diffusion layer (i.e. the wing area): $\lambda = -\langle\int_0^h dz \int_r^\infty \rho(x,y)dy\rangle \approx -hP_0S_0\int_r^\infty e^{-\kappa y}dy \approx hP_0S_0e^{-\kappa r}/\kappa$, where $h$ is the cell thickness, and $\langle\rangle$ represents averaging over the string length (x direction). Taking $P_0$~$5\times10^{-2}$ C/m$^2$, $S_0$=0.01, $r$ =0.2 μm, $1/\kappa$ =10 μm, $h$ =5 μm, we obtain $\lambda$~$5\times10^{-9}$ C/m which is the same order of magnitude as the value obtained from the electrophoretic experiments.

As the electrophoretic motion speed increases with the string length, a longer string defect can catch up and merge with a shorter one, as shown in Fig. S6. The dynamics of their merging involve the annihilation of the former -1/2 and the latter +1/2 topological defects, with two remaining half-integer defects becoming the head and tail of the merged string (SI Appendix, Video S2). The merged string length is equal to the sum of the lengths of the two individual strings before merging.

The interaction energy $F$ between two half-integer topological defects can be estimated from the Frank-Oseen energy in nematic liquid crystals and is typically a logarithmic function of their distance $R$, expressed as: $F = \pi K h \ln(R/r_c)/2$, where $K$ is the elastic constant in single $K$ approximation, $r_c$ is the core size of the topological defect (5). Since the string lengths remain unchanged during merging, there is no linear term that is present in the nematic phase (42). The interaction force between the +1/2 defect of the front string and the −1/2 defect of the rear string is given by: $f = dF/dR = \pi K h/2R$. The drag forces experienced by the front and rear strings are $F_{d1} = \zeta_1 v_1$ and $F_{d2} = \zeta_2 v_2$, respectively. The force balances for the front and rear string defects lead to $\zeta_1 v_1 = \rho E L_1 - \pi K h/2R$ and $\zeta_2 v_2 = \rho E L_2 + \pi K h/2R$, respectively. Using $dR/dt = v_2 - v_1$, we have:

$$\frac{dR}{dt} = \rho E\left(\frac{L_2}{\zeta_2} - \frac{L_1}{\zeta_1}\right) + \frac{\pi K h}{2R}\left(\frac{1}{\zeta_2} + \frac{1}{\zeta_1}\right) \tag{2}$$

Given that these two string defects have similar string lengths ($L_1 \approx L_2$), their viscous coefficients are approximately equal, $\zeta_1 \approx \zeta_2$, and we can obtain: $R = \sqrt{2\pi K h(t - t_0)/\zeta_1}$, with $t_0$ being a time constant.

The measured string separation $R$ within a 25 s time duration before the merging can be effectively fitted with this function, yielding the fitting parameter $Kh/\zeta_1$=0.25 μm$^2$/s (SI Appendix, Fig. S7). By considering $\zeta_1$~6×10$^{-6}$ N·s/m, we obtain the averaged elastic constant of the liquid crystal as $K$ = 1.5 pN, which is a reasonable value as compared with prior measurements (8, 43). In this analysis, we neglect the electrostatic interactions between the two string defects, considering the strong electrostatic screening by the counter ions.

**Effects of Different Ionic Doping and Proposed Charge Double Layer.** Based on the above experiments, we propose a charge double layer model for the string defects. The Gouy-Chapman diffusion layer in the butterfly wing regions is composed of the positive bound charge (i.e. $-\nabla \cdot \boldsymbol{P} > 0$) and the negative ions like $Cl^-$ and $Br^-$ dissociated from the polymers (Fig. 2E, Fig. 4A and 4B), and the Stern layer is composed of the positive charges carried by the string micelles and the bound charges (should be negative or $-\nabla \cdot \boldsymbol{P} < \boldsymbol{0}$) near the strings (Figs. 4A-B).

To fully understand the effects of ionic doping, we compared three distinctive types of ionic dopants in the RM734: cationic polymers, anionic polymers, and small molecular ionic compounds (Table 1). The presence of dopants with cationic polymer chains is essential to the formation of string defects in RM734 (Figs. 5A-C), either carbonate or siloxane chains yield similar results. Electrophoresis experiments can confirm that all string defects possess positive charges. To note, PDMOAP and poly(3-(trimethoxysilyl) propyl-trimethylammonium chloride) (PTTPA) that differ only by the alkyl chain (-$C_{17}H_{35}$) yield similar results, indicating that the role of steric effect is minor.

For anionic polymer dopants with ammonium and sodium ions bonded to polymeric carbonic and phosphoric chains, we observed no string defects but some chiral or complex domain structures (Figs. 5D-F). For small molecular ionic dopants such as sodium dodecyl sulfonate (SDS), octadecyltrimethylammonium chloride (OTAC), and cetyltrimethylammonium bromide (CTAB), we observed no string defects either but interesting textures that vary significantly among different dopants (Figs. 5G-I).

The fact that only cationic polymers can induce the formation of string defects implies that the interactions between the ionic dopants and the ferroelectric nematic liquid crystals are primarily of electrostatic nature, and that flexoelectric coupling also plays a role. Flexoelectric coupling refers to the coupling between electrical polarization and splay deformation(44, 45). The free energy associated with flexoelectric coupling can be expressed as $F_{flexo} = \int -\lambda \boldsymbol{P} \cdot \boldsymbol{n}(\nabla \cdot \boldsymbol{n}) d^3\mathrm{r}$, where $\lambda$ is the flexoelectric coupling coefficient (44, 45). Close packing of the pear-shaped RM734 molecules favours splay deformation, thus $\lambda$ is positive for RM734 molecules as indicated previous in theory and experiments (7, 46). Therefore, the flexoelectric coupling energy is minimized when the vector $\boldsymbol{P}$ and the splay $\boldsymbol{n}(\nabla \cdot \boldsymbol{n})$ are along the same direction.

In the presence of aggregates of doped polymers, close packing of the pear-shaped RM734 molecules around them favours that the splay $\boldsymbol{n}(\nabla \cdot \boldsymbol{n})$ points radially outward, and thus the polarization $\boldsymbol{P}$ around these string micelles should point outward as well to minimize the flexoelectric coupling energy. The outpointing $\boldsymbol{P}$ means that the bound charges near these strings are negative (i.e. $-\nabla \cdot \boldsymbol{P} < 0$), and thus only cationic polymers are effective in compensating these bound charges, consistent with our experimental results.

**Discussion and Conclusion**

To conclude, we show that ionic doping can enhance electrical screening in ferroelectric nematic liquid crystals and thus enables the formation of string defects with butterfly optical textures in polar fluids, consisting of half-integer topological defects paired by string polymeric micelles. We show that these string micelles are positively charged, and their wing regions exhibit positive bound charges. These findings support a charge double-layer model: the polymer chain micelles and densely packed RM734 molecules form a Stern layer, while small anionic ions and the positive bound charges in the wing regions constitute the diffusion layer. Importantly, only cationic doping effectively induces these unique string defects. We ascribe this finding to the positive flexoelectric coupling of the liquid crystal and the electrical screening provided by the doped molecules.

Recent study shows that pure nematic ferroelectric liquid crystals confined between glass plates exhibit chiral ground state to minimize electrostatic energy, a behaviour that can be significantly altered by ionic doping (26). The experiments presented here demonstrate that ionic doping can effectively eliminate the chiral state, resulting in a quasi-two-dimensional polarization configuration, and that cationic polymeric dopants are essential to balancing the bound charge induced by splay deformation. We anticipate that the topological configurations of string defects may be universal and could manifest in various forms. Future research could investigate how doping $N_F$ liquid crystals with rod-shaped or other geometrically structured colloidal particles influences the formation of topological defects and their interactions with bound charges and domain walls.

## Materials and Methods

**Ferroelectric liquid crystal and ionic doping materials.** The ferroelectric liquid crystal material RM734 was synthesized according to the procedures reported in previous publication (47), and purified by passing through chromatographic columns to remove stray ions. Ionic dopants dimethyl-octadecyl[3-(trimethoxysilyl) propyl] ammonium chloride (DMOAP, 60 wt% in methanol) was purchased from Shanghai Macklin Biochemical Co. Ltd.

The fluorescent ionic polymer dopants were synthesized by attaching the fluorescent molecules Rhodamine B with silane coupling agents (3-aminopropyl) triethoxysilane (3-APT), then hydrolysis in deionized water for 12 hours to form fluorescent-labelled silanol, and polymerization at 120°C for 30min. In experiments, the Rhodamine B (2 g, 2.1 mmol), N, N'-dicyclohexylcarbodiimide (DCC, 1 g, 2.5 mmol), 4-dimethyl aminopyridine (DMAP, 100.0 mg, 0.41 mmol) and 3-APT (1.2 mL, 2.5 mmol) were firstly dissolved in dichloromethane (DCM, 60 mL, 99%); the mixed solution was purged with pure nitrogen gas for ~20 minutes to remove oxygen gases, and then gently stirred at room temperature for 12 hours. The liquid products of this reaction were cleaned via suction filtration to remove precipitates of dicyclohexylurea, washed 3 times with hexanes to remove excess Rhodamine B monomers, and then vacuum-dried at room temperature to yield flaky purple solid product of Rhodamine-B-labelled 3-APT. The chemicals DCC, DMAP and DCM were purchased from Shanghai Aladdin chemical. Rhodamine B (99%) was purchased from Alfa Aesar.

**Processes for doping ionic dopants.** For dopants, including OTAC, CTAB, SDS, we firstly dissolved them in DCM and mix the resultant mixtures with a DCM solution of RM734 at a 1:2000 weight ratio of the dopant to RM734. For ammonium salt polymer ($NH_4PA$), ammonium polyphosphate ($NH_4PPA$), sodium polyacrylate (NaPA) and polyquaternium-28 (PN-28) we first dissolved them in DMSO for further dilution. We removed the solvent DCM by evaporation in vacuum at room temperature, to obtain RM734 with dopants as white solid.

For the polymeric dopants including DMOAP, 3-(trimethoxysilyl) propyl-trimethylammonium chloride (TTPA) and rhodamine b-amide trimethoxysilane, we dissolved in deionized water (2 wt%) for 2 hours firstly to form silanol via hydrolyzation. The solvent was removed by evaporation in vacuum at room temperature, yielding the solid of silanol oligomer. We firstly dissolved them in DCM and mix the resultant mixtures with a DCM solution of RM734 at a 1:2000 weight ratio of the dopant to RM734. We removed the solvent DCM by evaporation in vacuum at room temperature, to obtain RM734 with ionic dopants as white solid.

**Liquid crystal cell preparations.** We cleaned glass substrates with uniform ITO films or patterned ITO electrodes firstly with aqueous solutions of detergents and then with isopropanol in an ultrasonic bath. Then we blow-dried substrates by nitrogen gas and treated with oxygen plasma for 10 min to enhance wettability. We assembled a liquid crystal cell with two of these glass substrates with the cell gap controlled at 5 $\mu$m by using spherical glass spacers. We placed the RM734 powders as prepared above close to the cell gap and heated the cell to above the N-Iso transition temperature, causing automatic filling of the liquid crystal mixtures by capillary suction.

**Measurements of the polar director field.** We heated the liquid crystal cells to 150°C and then cooled down to and held at 125°C on a hot plate for 5 minutes to establish thermal equilibrium, and then measure the director fields around these string defects by the technique Pol-Scope imaging. To determine the polarization direction, we applied an in-plane electrical field (2 V/mm) and extract the polarization based on the response of the string defects.

**Confocal and SHG imaging of the string defects.** We firstly heated the liquid crystal cell to 150°C and cooling down to $N_F$ phase (125°C) by hot plate. To observe the string defects' structure at confocal microscope, we slowly cooled down (2°C/min) the liquid crystal cell to room temperature with glassy state which retained the string defects. To determine the distribution of the fluorescent group labelled cationic polymers, we used a confocal microscope (Zeiss, LSM980) to map the fluorescent intensity for string defect with a 40× water immersed objective. (excitation: 545-560 nm, emission: 570-580 nm).

**Acknowledgments**

Financial support by the National Key Research and Development Program of China via grant 2022YFA1405000, the National Natural Science Foundation of China via grant 6210030761, 12174177 and 12204226, and Guangdong Basic and Applied Basic Research Foundation via grant 2024B1515040023 are acknowledged.

## Figures and Tables

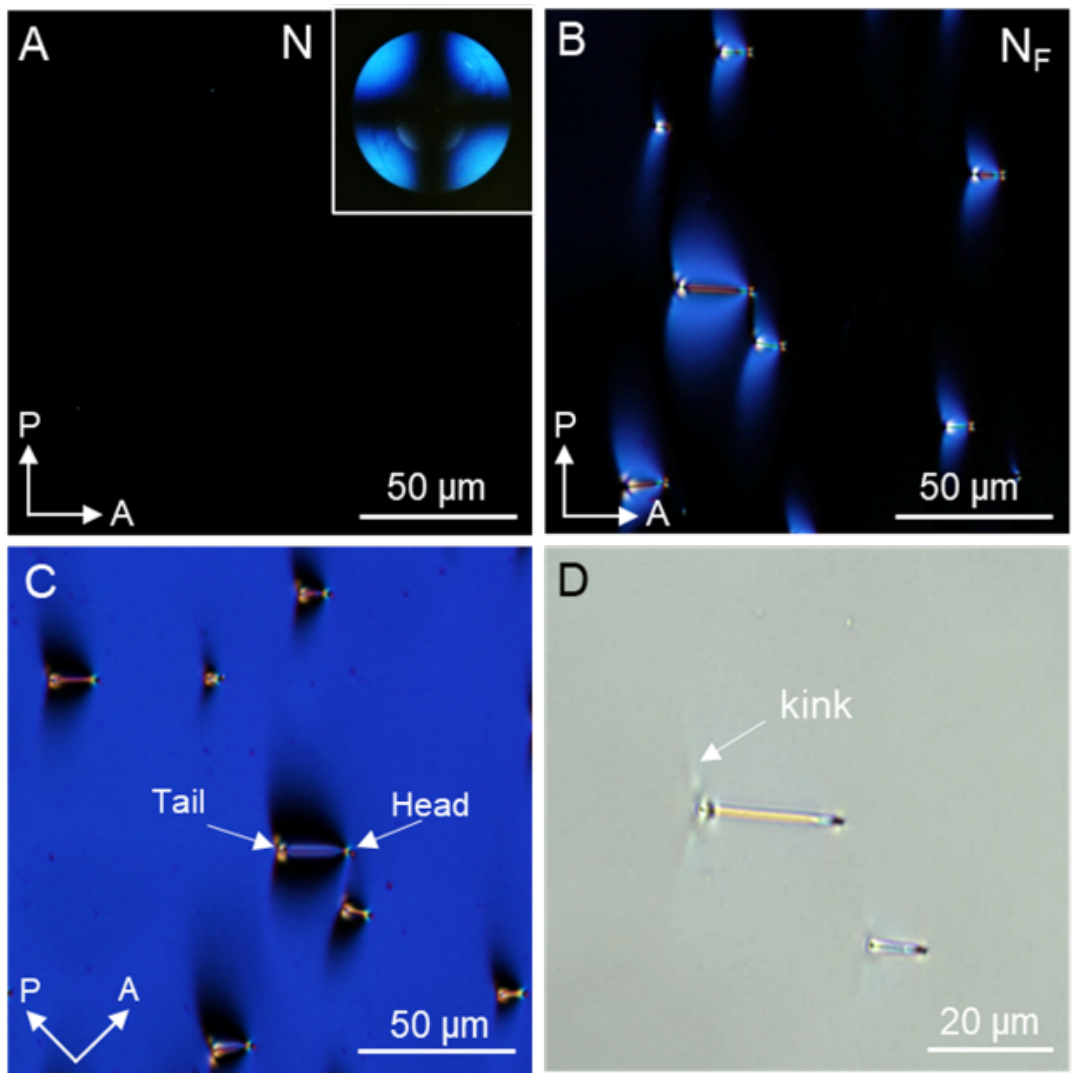

**Fig. 1. POM imaging of the string defects** (A) Representative cross-polarized optical microscopic image of a sample in the N phase at 140°C. The inset is a conoscopy image, indicating that the director is perpendicular to the plane. (B-D) Cross-polarized optical microscopic images (B, C), and a bright field image of string defects in the $N_F$ phase taken at 125°C (D), where the string defects are abundant and visible.

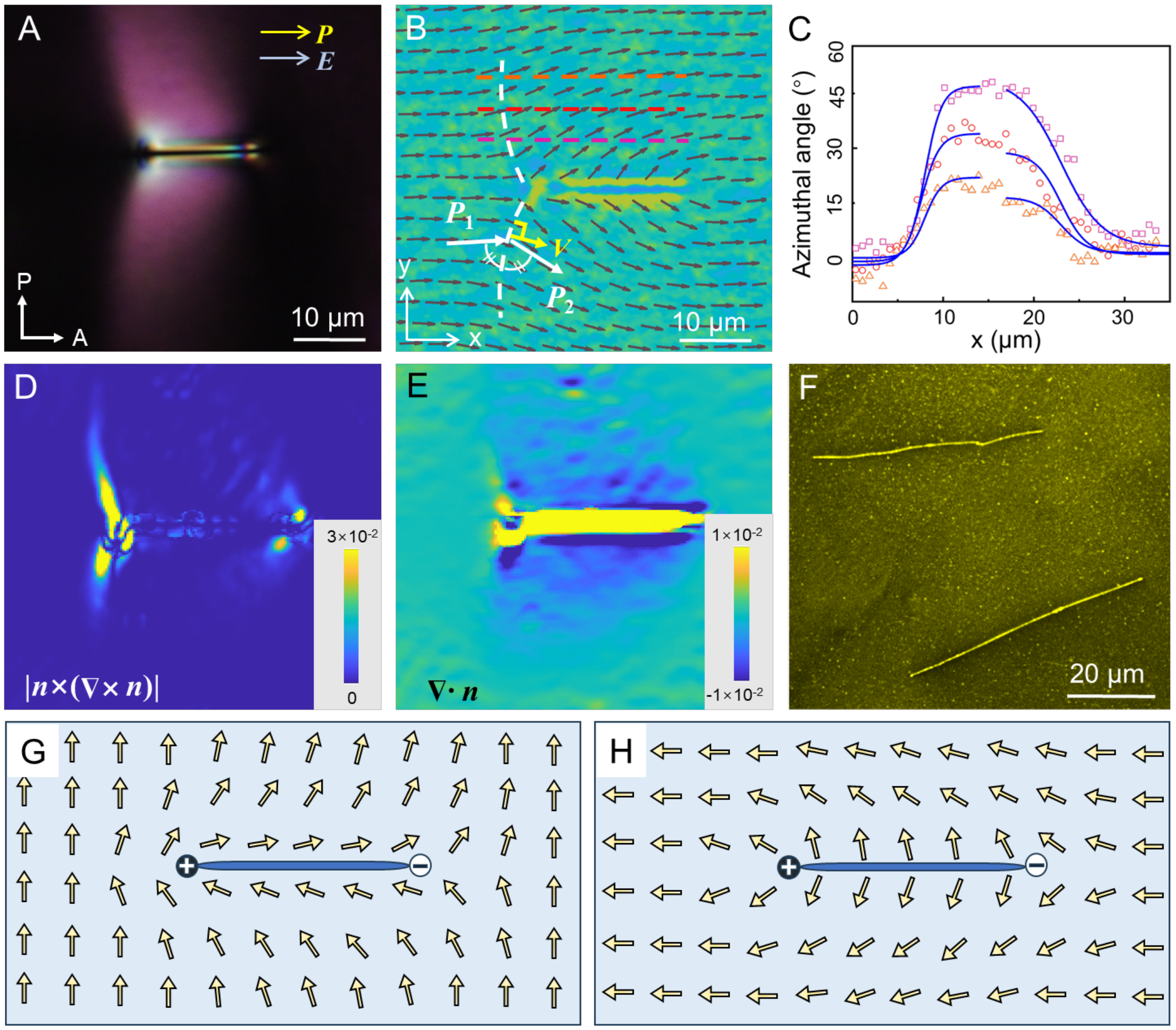

**Fig. 2. Optical characterizations of the string defects** (A) String defects under an in-plane DC electric field, showing that the polarization is parallel to the electric field. (B) Measured polarization field around a string defect, where the background colours represent the phase retardation. The white dashed lines represent the kink walls, which bisect the angle formed by the polarizations at two sides of them. (C) Representative profiles of the polarization angles with respect to the x-axis along the three dashed lines in (B). (D) Distribution of the bend deformations calculated form the polarization field in (B). (E) Distribution of the splay deformation calculated from the polarization field in (B). (F) A confocal microscopic image of two string defects for cationic polymer dopants labelled with Rhodamine B. (G-H) The schematic polarization field of a neutral string aligned perpendicular to neighbouring polarization field (G) and the schematic polarization field of a positive charged string aligned parallel to neighbouring polarization field (H).

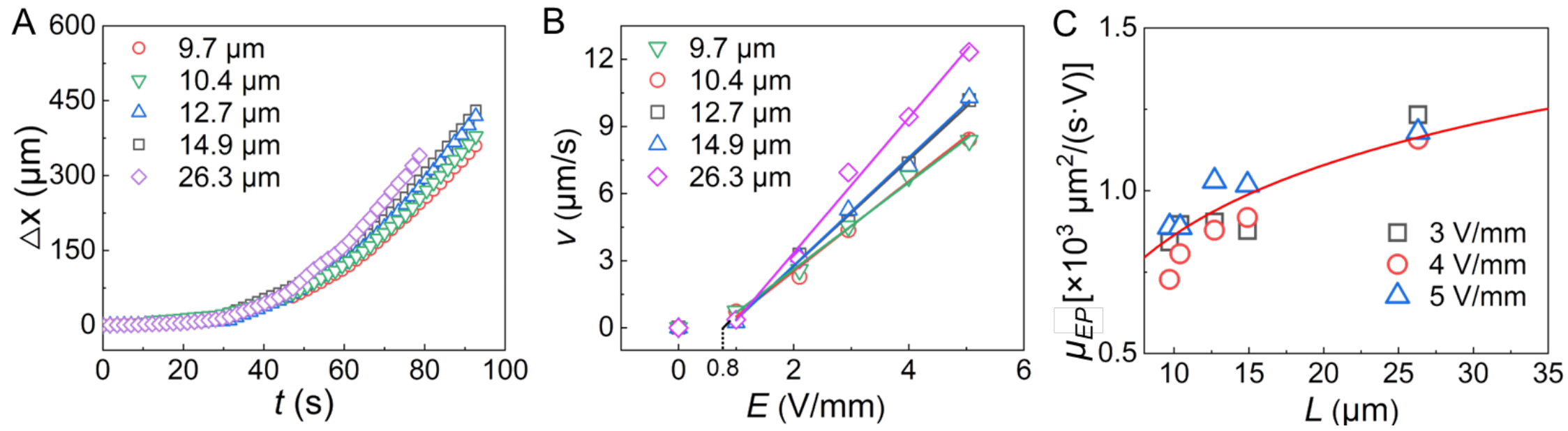

**Fig.3. Electrophoretic motion.** (A) Representative data of the measured x-coordinates of string defects as functions of time for different string lengths; (B) Measured motion velocity versus the electrical field strength for string defects with different lengths; (C) Dependence of the mobility $\mu_{EP}$ on the string length measured at different electric field strengths. The red curve is the fitting curve using the theoretical formula for $\mu_{EP}$.

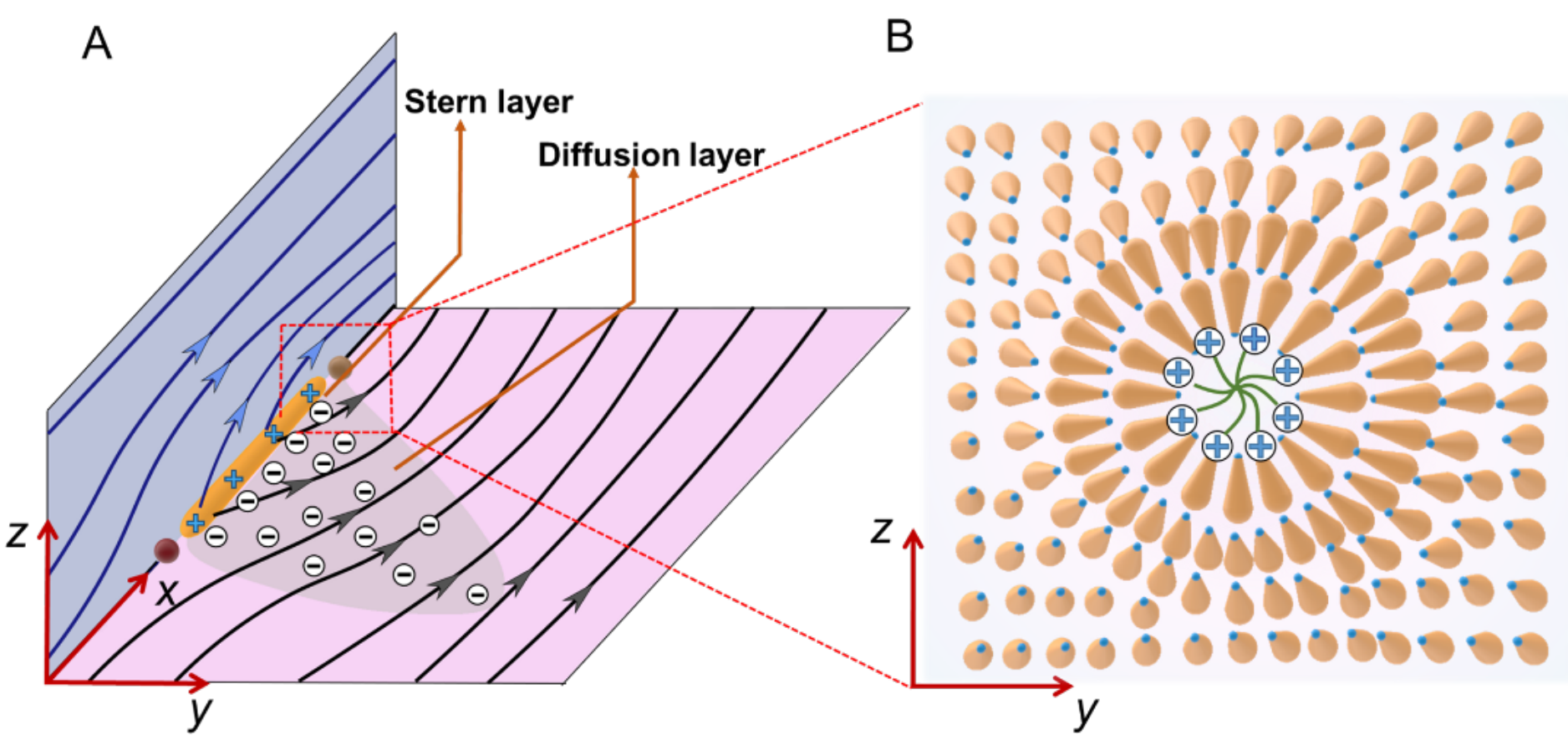

**Fig. 4. Schematic of RM734, polarization field and double charge layers of the string defects.** (A) The polarization field and distribution of the doped ions in the xz and xy planes of a string defect. The anionic ions and the positive bound charges (not shown) due to the splay deformations form the Gouy-Chapman diffusion layer. (B) The cross section of a string defect in the yz plane (red dash line in A). The close-packing of the RM734 molecules leads to outward pointing of the electrical dipoles and thus negative bound charges. The strings made of positively charged polymers form a stern charge layer.

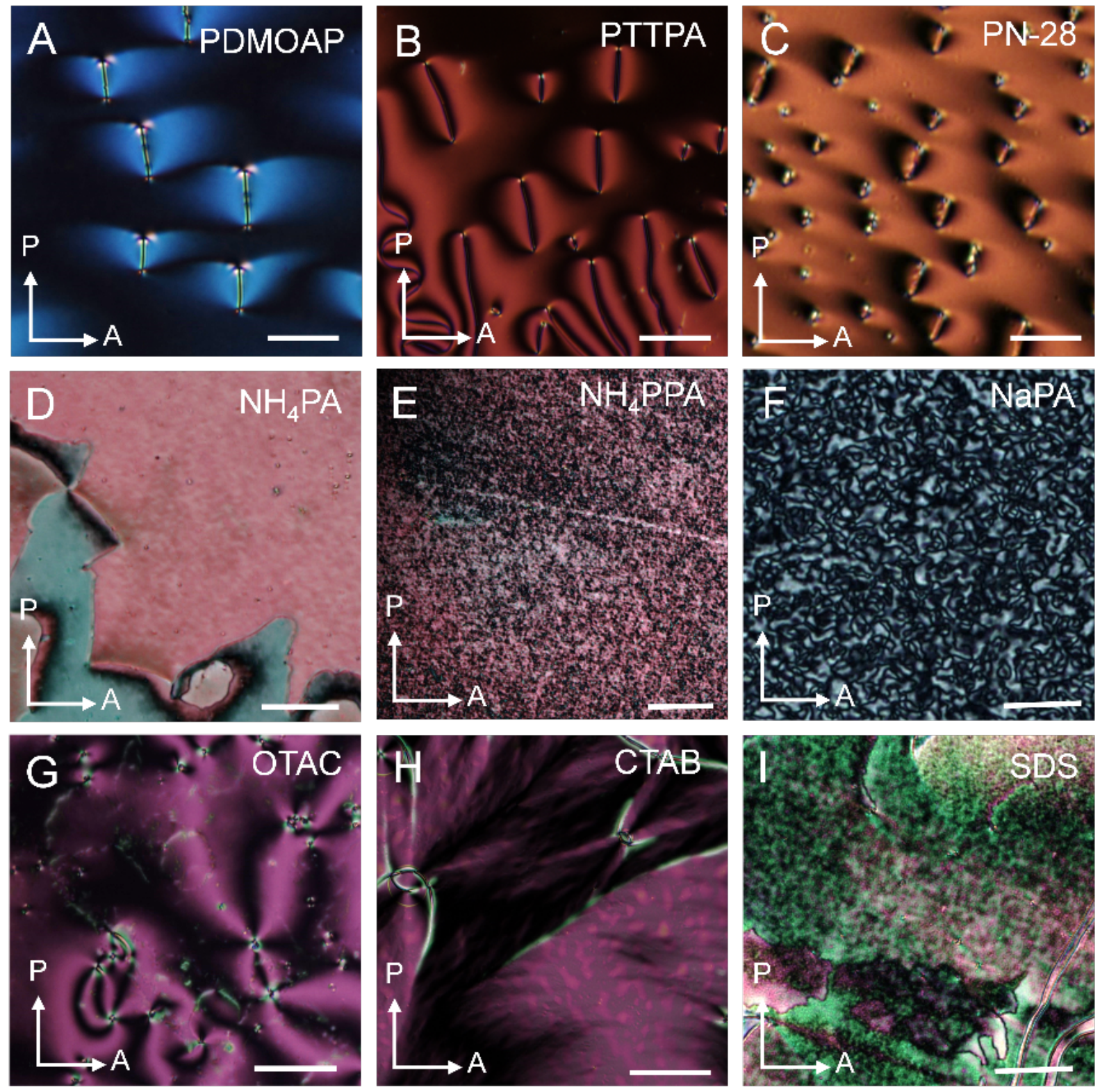

**Figure 5. Effects of different ionic doping molecules.** (A-C) Textures of the RM734 $N_F$ phase doped with cationic polymers PDMOAP (A), PTTPA (B), and PN-28 (C); (D-F) Textures of the RM734 $N_F$ phase doped with anionic polymers $NH_4PA$ (D), $NH_4PPA$ (E), and NaPA (F); (G-I) Textures of the RM734 $N_F$ phase doped with ionic compounds OTAC (G), CTAB (H), and SDS (I). All the texture images were taken at temperature 120°C, with doping concentrations all at 0.05 wt%. Scale bars are 20 μm in all images.

**Table 1.** Dopants with different chemical structures

| Ionic type | Chemicals | Chemical Structure | Formation of String Defects | String Charge |
|---|---|---|---|---|
| | PDMOAP | $C_{18}H_{37}$ Cl- $N^+$ O Si n O | Yes | Positive |
| **Cationic polymer** | PTTPA | Cl- $N^+$ O Si n O | Yes | Positive |
| | PN-28 | n m O NH N O $N^+$ Cl- | Yes | Positive |
| | $NH_4PA$ | n $NH_4^+$ $^-O$ O | No | |
| **Anionic polymer** | $NH_4PPA$ | O O P n $O^-$ $NH_4^+$ $O^-$ | No | |
| | NaPA | n $Na^+$ $^-O$ O | No | |
| | OTAC | $C_{18}H_{37}$ $Cl^-$ $N^+$ | No | |
| **Ionic compound** | CTAB | $C_{16}H_{33}$ $Br^-$ $N^+$ | No | |
| | SDS | $C_{12}H_{25}$ O O S O $Na^+$ $^-O$ | No | |